\documentclass{emulateapj}
\usepackage[leftcaption]{sidecap}
\usepackage{microtype}
\usepackage{graphicx,natbib}
\def\cK{{\cal K}} 
\newcommand{\mps}{m\,s$^{-1}$}

\shorttitle{Inversions for Average Supergranular Flows}
\shortauthors{Michal \v{S}vanda}

\begin{document}

\title{Inversions for Average Supergranular Flows Using Finite-frequency Kernels}
\author{Michal \v{S}vanda}
\affil{Astronomical Institute, Academy of Sciences of the Czech Republic (v. v. i.), Fri\v{c}ova 298, CZ-25165 Ond\v{r}ejov, Czech Republic}
\affil{Charles University in Prague, Faculty of Mathematics and Physics, Astronomical Institute, V Hole\v{s}ovi\v{c}k\'ach 2, CZ-18000 Prague 8, Czech Republic}
\email{michal@astronomie.cz}

\begin{abstract}
I analyse the maps recording the travel-time shifts caused by averaged plasma anomalies under an ``average supergranule'', constructed by means of statistical 
averaging over 5582 individual supergranules with large divergence signals detected in two months of HMI Dopplergrams. By utilising a three-dimensional validated time--distance inversion code, 
I measure the peak vertical velocity of 117$\pm$2~\mps{} in depths around 1.2~Mm in the centre of the supergranule and root-mean-square vertical velocity of 21~\mps{} over the area of the supergranule. 
A discrepancy between this measurement and the measured surface vertical velocity (a few \mps) can be explained by the existence of the large-amplitude vertical flow under the surface of supergranules with large divergence signals,
recently suggested by \cite{2012SoPh..tmp..136D}.
\end{abstract}

\section{Large-magnitude subsurface supergranular flows?}
The nature of supergranules -- convection-like structures observed in 
the solar photosphere -- is largely still in debates \citep[see a review by][]{lrsp-2010-2}. Only the surface properties of supergranules are well established. 
The plasma flows within the supergranules are predominantly horizontal with root-mean-square velocity  
$\sim$300~\mps. The vertical component of the supergranular flow is difficult 
to measure, its amplitude is usually within measurement error 
bars. Statistically it has been determined that the root-mean-square vertical velocity ranges from 4~\mps{} \citep{2010ApJ...725L..47D}
to 29~\mps{} \citep{2002SoPh..205...25H}. 

The deep structure of supergranules is practically unknown. Attempts were made using local helioseismic methods
with controversial results. Some studies \citep[e.g.][to name a few]{1998ESASP.418..581D,2003ESASP.517..417Z} 
revealed supergranules as convection-like cells extending to depths of 8--25~Mm with a deep ``return flow''. \cite{2007ApJ...668.1189W} and \cite{2008SoPh..251..381J} did not detect the flow reversal, but pointed out that any inversion for the flow snapshot deeper than 4--6~Mm is dominated by the random noise and thus does not reveal any information about the deep supergranular flow. \cite{2012ApJ...749L..13H} recently showed that supergranules may extend to depths equal to their widths indicating that analysis of supergranules with differing sizes may lead to a different depth structure. Numerical simulations of near-surface Sun-like convection \citep[e.g.][]{2008ASPC..383...43U} indicated that the amplitude of the flows on 
supergranular scales decreased with increasing depth, although the flows were highly structured on smaller scales. 

Recently, a very 
surprising result came to light from helioseismology -- 
\cite{2012SoPh..tmp..136D} claimed that the careful analysis of travel-time maps provided evidence for a large-amplitude flow under the surface of supergranules. No 
inversion was performed in this work. The authors detected a systematic offset in the travel-time shifts measured for large 
separations between the measurement points using a special spatio-temporal filtering of the data. This 
non-standard procedure was selected to avoid the cross-talk between the horizontal and vertical 
flow in supergranules, which was known to be hard to avoid in inverse methods. They concluded that the offset was due to the large-amplitude vertical flow below the surface. Using a 
simple Gaussian model, constrained by surface measurements, they predicted a peak in the 
vertical flow of 240~\mps{} at a depth of $2.3$~Mm.  

In this study, I use a recently implemented, improved 
time--distance inversion code (which minimises the cross-talk 
between the flow components, the main issue leading \citeauthor{2012SoPh..tmp..136D} to use non-standard time--distance methods) to verify their hypothesis.

\begin{SCfigure*}[\sidecaptionrelwidth][!th]
\centering
\rule{0.5cm}{0pt}\includegraphics[width=0.6\textwidth]{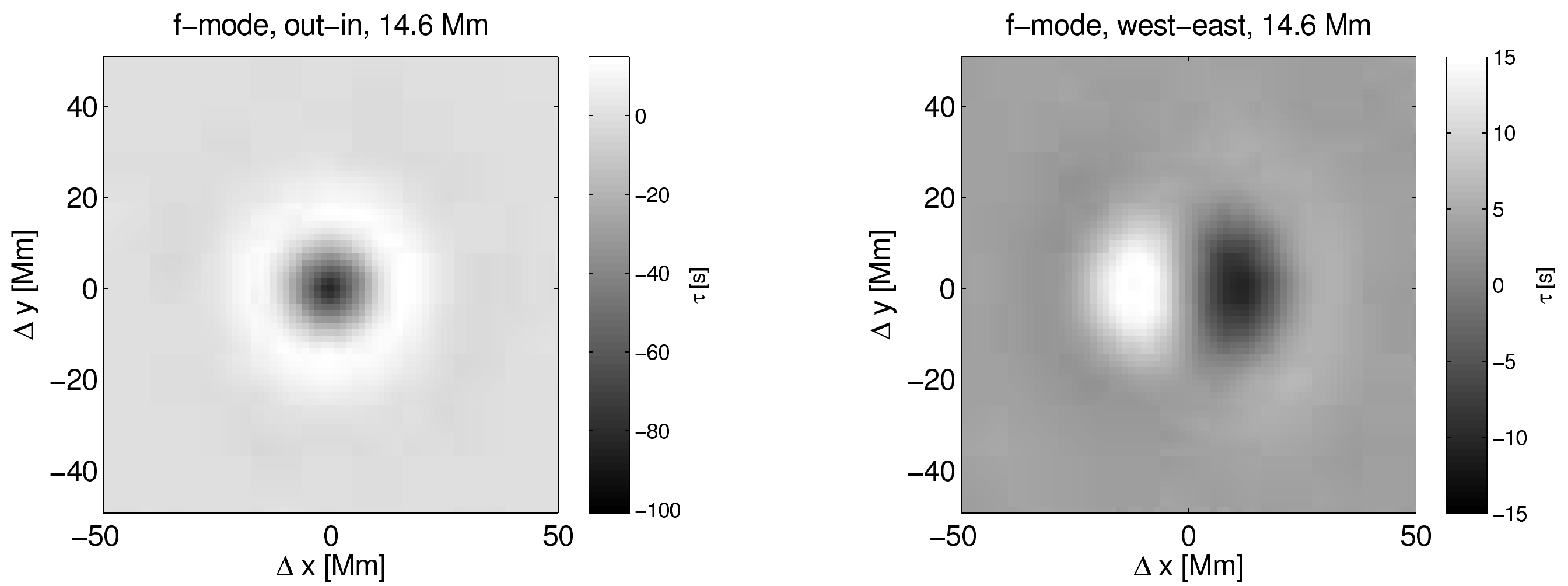}
\caption{Two examples of average difference travel-time maps used in the 
inversion. The $f$-mode travel times with centre-to-annulus and 
centre-to-quadrant geometry with annulus radius 14.6~Mm are presented. 
The out--in measurement is sensitive to the vertical flow and the 
divergence of the horizontal flow, while west--east measurement is 
practically sensitive to the west--east component ($v_x$) of the 
horizontal flow only. }
\label{fig:ttimes}
\end{SCfigure*}

\section{Inversion setup}
\label{sect:inversion}
The code used in this study (described and tested in \citealt{Svanda2011}) implements Subtractive Optimally 
Localised Averaging \citep[SOLA;][]{1992AA...262L..33P} three-dimensional time--distance \citep{1993Natur.362..430D} inversion. The SOLA algorithm allows to balance between 
various terms in the cost function. Among those, important 
terms evaluate the localisation in the Sun (an averaging kernel), the level of random noise in the 
results, and cross-talk contributions. By choosing values of the trade-off parameters, one can 
regularise strongly about some of these terms, while relaxing demands on the others.

The properties of the hypothetical large-amplitude supergranular flow require a special setup of the 
time--distance inversion. To avoid misleading signals from the side-lobes of the averaging kernels, 
one has to pick the solution with good localisation and therefore higher noise level. Luckily, one 
can use the statistical approach -- the trade-off parameter is setup so that the averaging kernel 
is highly localised, while the noise level is too large (of the order of 100~\mps) to do the tomography of individual supergranules. 
By averaging $N$ supergranules (containing independent realisation of the travel-time noise) the noise level is reduced $\sqrt{N}$-times. 

The SOLA inversion was performed in a Cartesian approximation ($x$ and $y$ for the horizontal 
coordinates and $z$ for the vertical) using ridge-filtered sensitivity 
kernels (modes $f$ to $p_2$) with centre-to-annulus and 
centre-to-quadrant averaging geometries \citep{1997ASSL..225..241K}. The radii of the 
annuli ranged from 5 to 20 pixels, with a pixel size of $1.46$~Mm ($0.12^\circ$). The ridge filtered sensitivity 
kernels were computed using the Born approximation \citep{2007AN....328..228B}. Only the near sub-surface layers were targeted.

\subsection{Travel-times}

In the current study, I used data similar to \cite{2012SoPh..tmp..136D}. Dopplergrams from 
Helioseismic and Magnetic Imager (HMI; \citealt{2012SoPh..275..207S}, 
\citealt{2012SoPh..275..229S}) were used to 
produce 64 consecutive twelve-hour datacubes for the travel-time measurements, starting on 10 June 
2011. The inversion was designed to be applied to travel times measured from 
Michelson Doppler Imager (MDI; \citealt{1995SoPh..162..129S}) full-disc Dopplergrams. Therefore, the HMI Dopplergrams were degraded to MDI full-disc 
resolution by means of down-sampling and application of a wave-vector filter, so that the power spectrum of the 
degraded HMI datacubes was close to the power spectrum of MDI full-disc datacubes. Only the disc-centre region was tracked in each datacube; the tracking and mapping was done using the code 
{\sc drms\_tracking} (Schunker \& Burston, 
unpublished) implemented in German Science Center for SDO. The Dopplergram datacubes were spatio-temporally filtered with ridge filters, separating $f$, $p_1$, and $p_2$ modes. The travel times were measured following \cite{2004ApJ...614..472G}, using centre-to-annulus and 
centre-to-quadrant geometries with radii of the 
annuli 5 to 20 pixels (consistently with the sensitivity kernels used in the inverse problem). 
\section{An average supergranule}

To reduce the level of random noise in the flow maps retrieved by good-localisation inversion, I constructed a set of travel-time maps representing an `average 
supergranule', thus reducing the level of random noise in travel times. 

\subsection{Identification of supergranules}

The locations of individual supergranules were determined using an algorithm similar to that used by
\cite{2010ApJ...725L..47D} and \cite{2012SoPh..tmp..136D}. Supergranules were searched for in 
the centre-to-annulus difference travel-time maps measured from $f$-mode filtered datacubes for the 
distance range of 8--12 pixels (0.96--1.44$^\circ$). The travel-time maps were smoothed with a 
Gaussian window with Full-Width-at-Half-Maximum of 3~Mm to suppress random fluctuations. Such maps were sensitive to the horizontal 
divergence of the flow assumed to occur inside supergranular cells.

\begin{enumerate}
\item All pixels, where the sound travel time was smaller than the mean by more than two standard deviations, were tagged. This marked the regions with large divergence signal, where the locations of the 
supergranular cell centres could be expected. 
\item For each tagged point a square neighbourhood (with a side length of 30~Mm) was taken and the position of the maximal divergence (thus minimal travel time) in this small 
region was determined. The location of this pixel was tagged. As a consequence, several points were possibly
tagged in the large-divergence region of the same supergranule.
\item All tagged points were cycled and the distance to all other tagged points was calculated. 
Should the distance between any pair of points be less than 23~Mm, only the point with the 
minimal value of the travel-time was preserved. Only one point  
then represented the original cluster of points. This step removed duplicities. 
\end{enumerate}

\noindent In all 64 datacubes, the procedure identified 5582 supergranule centres.
\begin{SCfigure*}[\sidecaptionrelwidth][!th]
\rule{0.5cm}{0pt}\parbox{0.6\textwidth}{\includegraphics[width=0.6\textwidth]{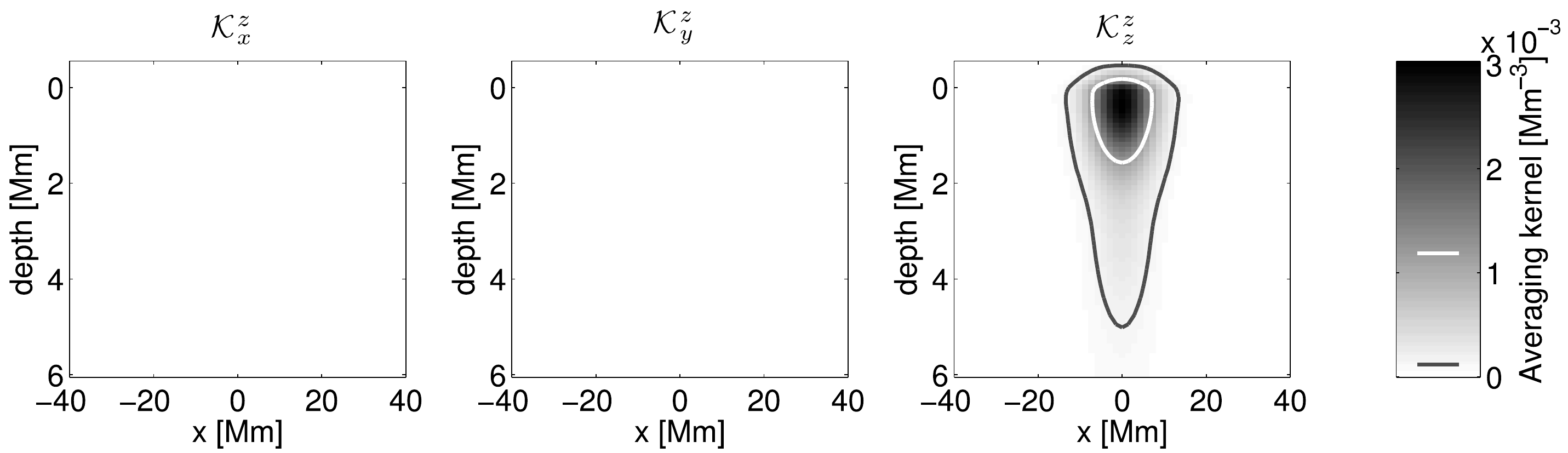}\\
\rule{0.6\textwidth}{1pt}\\
\includegraphics[width=0.6\textwidth]{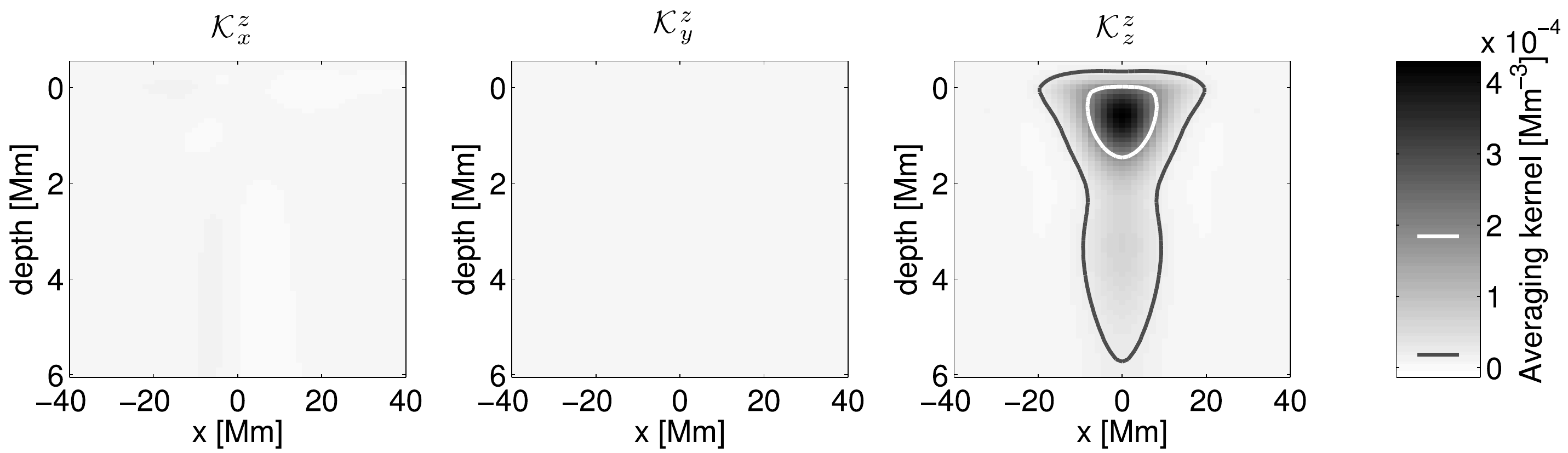}}
\caption{Vertical cuts through averaging kernels for the good-localisation (upper) and low-noise (lower) inversion 
of near-subsurface vertical flows. Horizontally, the averaging kernels are roundish. Over-plotted 
contours, which are also marked on the colour bar for reference, denote 
the following: half-maximum of the kernel (white) and $5$\% of the 
maximum value of the kernel (grey). Important component of the averaging kernel ($\cK^z_z$) is better localised in near sub-surface layers in the case of good-localisation inversion. Noise levels of these inversions are discussed in the main text. It is also evident that the contribution of the cross-talk is negligible as the 
cross-talk averaging kernels ($\cK^z_x$ 
and $\cK^z_y$) are practically zeroes. }
\label{fig:akerns}
\end{SCfigure*}

\subsection{Near-subsurface flows}
For the inverse modelling, the set of all 64 travel-time maps was averaged about the location of 
the supergranular centres. This resulted in a single set of travel-time maps, representing travel-time shifts 
caused by averaged anomalies under the average supergranule. Examples are 
displayed in Fig.~\ref{fig:ttimes}. 

The averaged travel-time maps were independently inverted for all flow components. The averaging of travel-time maps over 5582 supergranules 
reduced the noise level by a factor of $\sqrt{5582}=76$, thus having the level of random noise in the inverted velocities less 
than 2~\mps{}. Averaging kernels for the vertical component of the flow are displayed in 
Fig.~\ref{fig:akerns} (upper panel). The flow estimates are averaged mostly over depths of 
0--2~Mm with some contribution from deeper layers. The centre of gravity of the averaging kernels is located at a depth of 1.2~Mm. 

\begin{figure}[!b]
\centering
\includegraphics[width=0.35\textwidth]{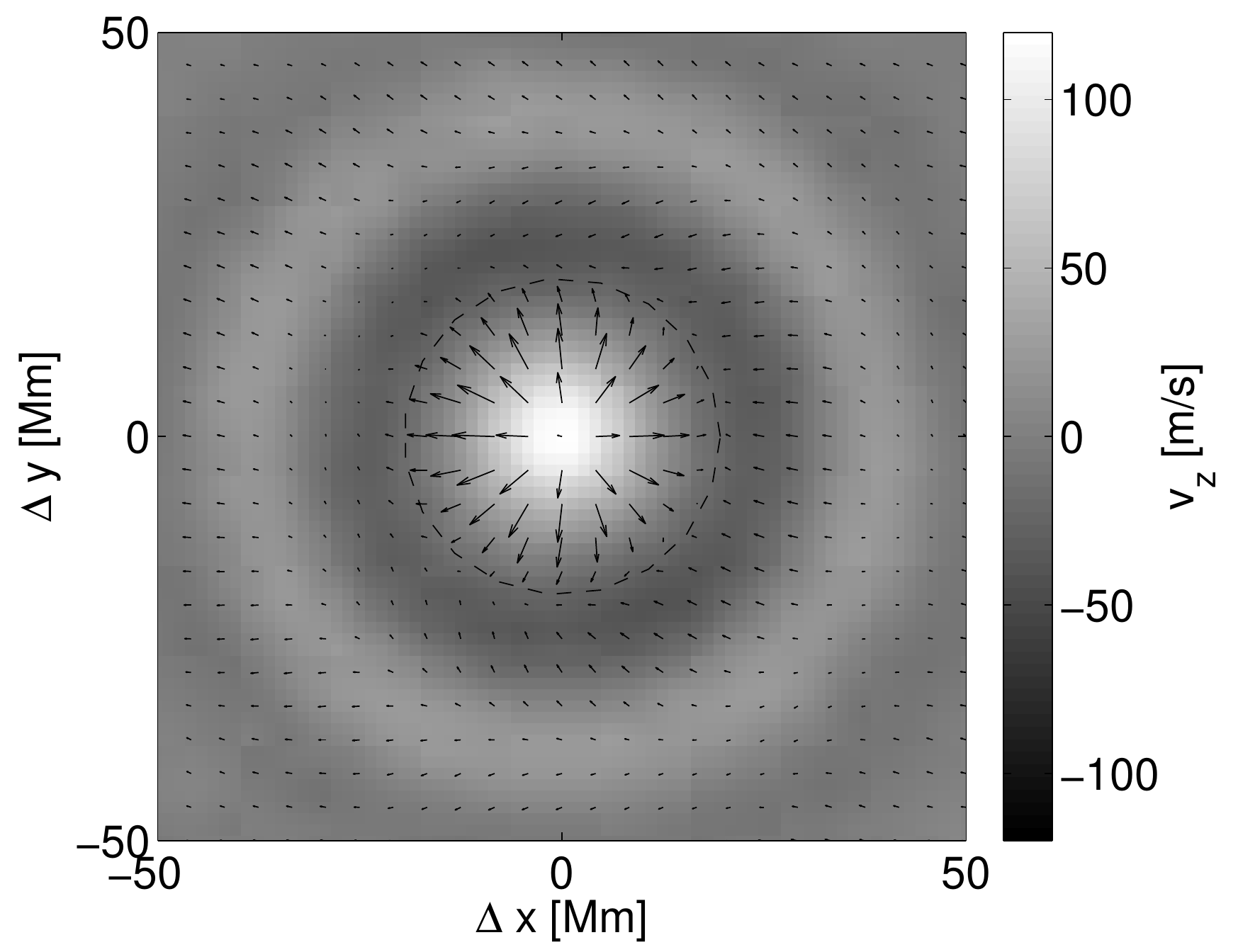} 
\caption{The flow field in the 
average supergranule at depths around 1.2~Mm. Vertical component ($v_z$) is indicated by 
colours, horizontal component by arrows. The largest arrow corresponds to the 
flow of 250~\mps. The large annulus of positive vertical velocity 
denotes the average location of the neighbouring supergranules. A 
dashed line indicates the half-way to the neighbouring supergranules, 
thus indicating the size of the average supergranule.}
\label{fig:avgsg}
\end{figure}

The resulting flow field in the average supergranule at depths of around 1.2~Mm is 
displayed in Fig.~\ref{fig:avgsg}. While the horizontal components of the flow velocity  
agree with what is expected from the literature, the magnitude of the vertical flow is very large, 
in agreement with \cite{2012SoPh..tmp..136D}. Note that the inverted flow roughly fulfills mass conservation. 

\begin{figure}[!t]
\centering
\includegraphics[width=0.35\textwidth]{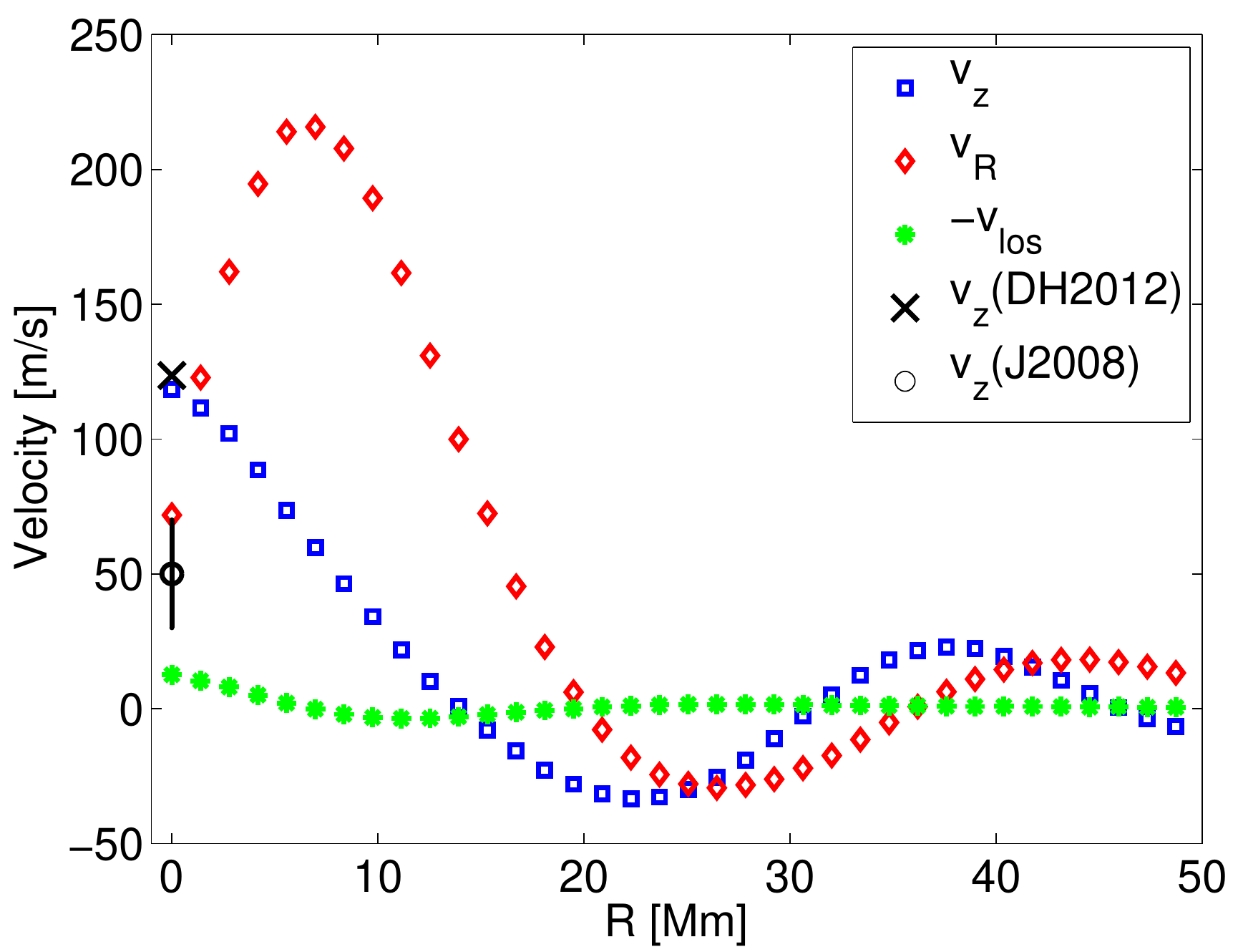}
\caption{Azimuthal average of the radial (with respect to the cell centre, 
red diamonds) and vertical ($v_z$, blue squares) near-subsurface velocity in the average 
supergranule. The line-of-sight velocity ($v_{\rm los}$ with negative 
sign; green full circles) obtained from the stacked Dopplergrams of supergranules 
is over-plotted for reference. The green curve is practically identical 
to that presented in \cite{2010ApJ...725L..47D}. Error bars for all plots are 2~\mps{} or less. Results from other studies are also overplotted for reference: the best-fit model of \cite{2012SoPh..tmp..136D} at the depth of 1.2~Mm is with the black $\times$ and the maximum range of the vertical flows measured in supergranular centres at the depth of $\sim1$~Mm by \cite{2008SoPh..251..381J} with the black open circle and error bar. Note that at the depth of 1~Mm the best-fit model of \cite{2012SoPh..tmp..136D} predicts the vertical flow of 76~\mps. }
\label{fig:radialflows}
\end{figure}

The magnitude of the vertical and horizontal flow, as a function of the distance $R$ from the centre 
of the average supergranule, is plotted in Fig.~\ref{fig:radialflows}. The vertical velocity peaks in the cell 
centre with a value of 117~\mps. The root-mean-square vertical velocity over the area of the average supergranule is 21~\mps, 
well within the measurement by \cite{2002SoPh..205...25H}. The 
second peak in the vertical velocity, at a distance of 38~Mm from the cell centre, indicates the 
average location of the neighbouring supergranules in my sample. 

The known location of the supergranules allows one to also average the measured Dopplergrams and obtain the average line-of-sight velocity, which corresponds to the vertical 
velocity at the surface. My results are similar to those of 
\cite{2010ApJ...725L..47D}, i.e. 12$\pm$1~\mps{} in the cell centre.  

\subsection{Comparison to regular sub-surface flow field}
\begin{figure}[!t]
\centering
\includegraphics[width=0.35\textwidth]{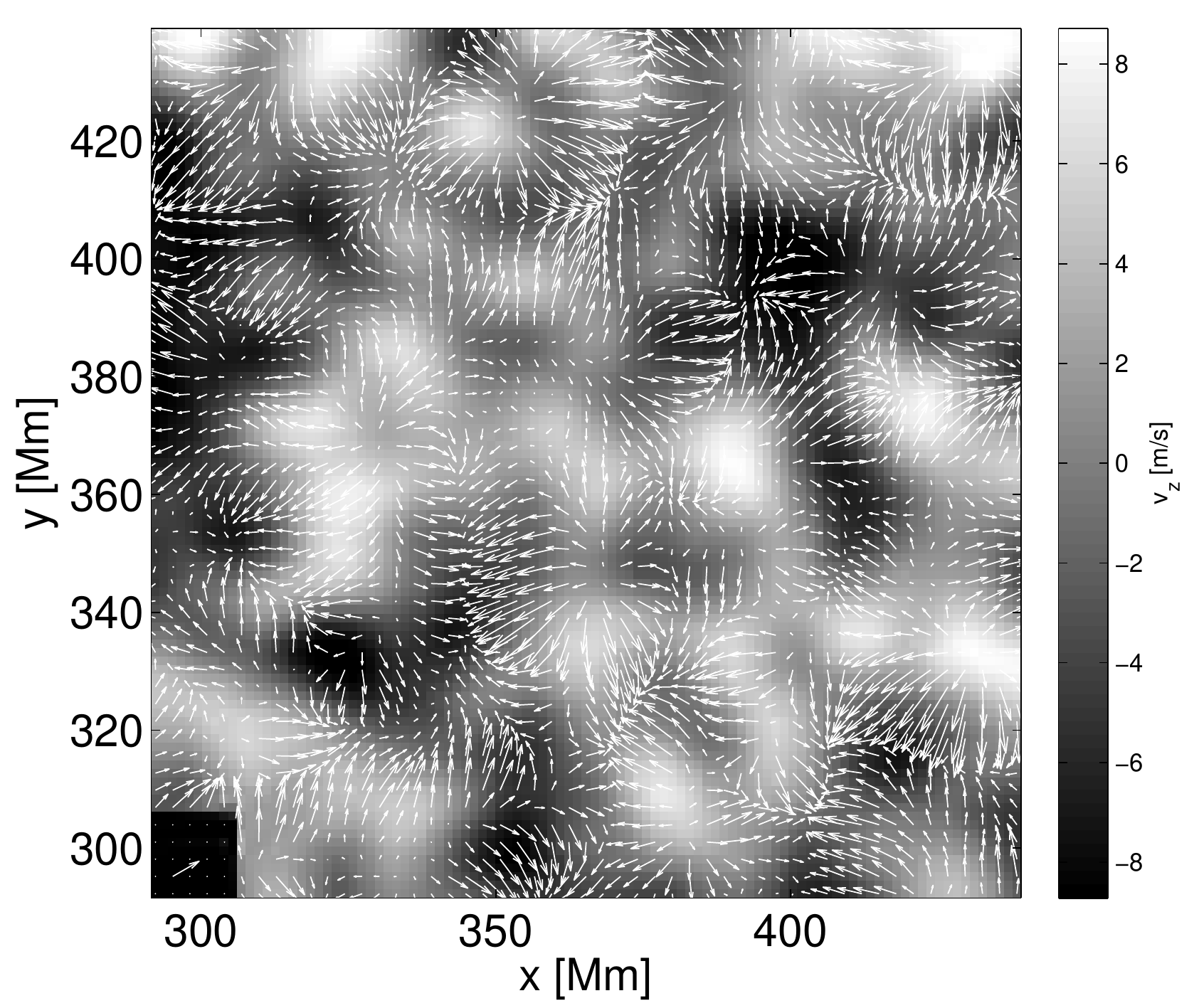}
\caption{An example of the near-subsurface flow field obtained by the low-noise inversion. The measurement were averaged 
over 1~day, the predicted errors are 34~\mps{} in horizontal 
components and 5~\mps{} in vertical component. The scale arrow denotes the horizontal flow of 250~\mps.}
\label{fig:surfaceflow}
\end{figure}

To check the consistency of my inverse modelling, I ran another inversion with the same setup and the same target depth, only regularising more strongly about the random-noise level term. This approach allows one to ``see'' a snapshot of the near sub-surface flow field on supergranular scales. An example flow map achieved with this approach is displayed in Fig.~\ref{fig:surfaceflow}. Obviously, no 100-\mps{} vertical flows are present in such a map with vertical velocity peaking at around 10~\mps{}, well within the limits established by previous studies. Regularisation about the noise term in the inversion cost function leads to the poorer localisation in the Sun, in particular leading to wide lobes in the averaging kernel (see Fig.~\ref{fig:akerns}). Application of the low-noise inversion to the average-supergranule travel-time 
maps provides a peak vertical velocity of 5~\mps{}, twenty-time less than the application of the good-localisation inversion. 

The obvious discrepancy can be explained by the differences in averaging kernels. To verify this hypothesis, I construct a simple model of the supergranular vertical flow, following both \cite{2010ApJ...725L..47D} and \cite{2012SoPh..tmp..136D}. The supergranular vertical flow field $v_z	(R,z)$ is modelled as
\begin{equation}
v_z(R,z)=v_0 {\rm J}_0 \left(2\pi \frac{R}{3L}\right)   \exp{\left[ -\frac{(z-z_0)^2}{2\sigma_z} \right]} ,
\end{equation}
where the horizontal part is approximated by a Bessel function of the first kind with $L=15.1$~Mm being the half-size of the supergranule adopted from \cite{2010ApJ...725L..47D}, and the overall magnitude is scaled by a Gaussian vertical envelope with $v_0=240$~\mps, $z_0=-2.3$~Mm, and $\sigma_z=0.912$~Mm, following the simple model used by \cite{2012SoPh..tmp..136D}. In this approximation, the vertical flow peaks at the value of 240~\mps{} at the depth of $2.3$~Mm and decreases towards the surface, with a magnitude of 10~\mps{} at surface levels. 

I convolve the model for the vertical flow with the averaging kernels resulting from both the good-localisation and low-noise inversions. The resulting maps do not contain any realisation of the random noise, but, due to the expected noise levels less than 2~\mps, it can be considered negligible. The resulting peak vertical velocities that are retrieved by the inversions from the modelled flow field are $26$~\mps{} in for case of the good-localisation inversion, and $2.8$~\mps{} for the case of the low-noise inversion. 

The values retrieved from the model do not agree with the values retrieved from the travel-times inversion. There is one order of magnitude difference which can sufficiently be explained by the different averaging kernels. This order of magnitude difference agrees well with the order of magnitude difference obtained from the average-supergranule travel-time maps. This naive experiment demonstrates that the existing large-amplitude flow might be hidden for some inversions with extended averaging kernels. 
\section{Concluding remarks}

The very large amplitude of the vertical flow measured by time--distance 
helioseismology is somewhat surprising. Not indicating possible problems in the forward and inverse modelling, there are 
three main issues that could lead to biassed results. 

\begin{itemize}
\item{\bf Identification of supergranules} The method used here prefers the supergranules with highest divergence signals. It 
detected only 5582, while something like 50000 is expected, when approximating the supergranule 
by the cell with characteristic size 30~Mm in a hexagonal pattern covering the whole Sun. Thus, the use of a better segmentation algorithm to identify all cells may bring a new insight into the problem.
\item{\bf Stacking of supergranules} The detected supergranules are stacked 
to the point of the maximal divergence, which does not necessary need 
to be in the geometrical centre. To overcome this issue, I modified 
the detection code to fit paraboloids in the vicinity 
of the points with the highest divergence signals and to choose vertices of these paraboloids as 
locations 
of the cell centres. This did not change the results at all. 
\item{\bf Foreshortening} In this study, all safely detected supergranules 
lying within 24 degrees from the disc centre were used to stack 
the average supergranule. In the sensitivity kernels, foreshortening 
is not accounted for. Limiting the location of supergranules used  
of 15 degrees in heliocentric angle, only 1014 supergranules were 
detected. The amplitude of the vertical velocity did not change at all.
\end{itemize}

\noindent I conclude that there are supergranules on the Sun (around 10\% of the expected count of all supergranules) which exhibit large divergence signals connected with the large-amplitude subsurface flows. These flows are probably very narrow and might have been missed in the past by inversions with a low noise level preference. The model of the ``average supergranule'' inferred here is representative only to this small fraction of all expected supergranules and does not have to describe the true representative cell. To do so, a better segmentation method must be used to construct a model of a ``true average supergranule''.  

\acknowledgements I acknowledge the support of the Czech Science 
Foundation (grant P209/12/P568) and of the Grant Agency of Academy of 
Sciences of the Czech Republic (grant IAA30030808). This work utilised the resources and helioseismic products 
dispatched within the German Science Center at MPS in Katlenburg-Lindau, Germany, which is supported by German 
Aerospace Center (DLR). The data were kindly provided by the HMI consortium. The HMI project is supported by NASA contract NAS5-02139.
%

\end{document}